\documentclass[fleqn,usenatbib]{mnras}

\usepackage{newtxtext,newtxmath}

\usepackage[T1]{fontenc}

\DeclareRobustCommand{\VAN}[3]{#2}
\let\VANthebibliography\thebibliography
\def\thebibliography{\DeclareRobustCommand{\VAN}[3]{##3}\VANthebibliography}

\usepackage{graphicx}	
\usepackage{amsmath}		
\usepackage{comment}
\usepackage[normalem]{ulem}

\title[Influence of adiabatic heating on PUIs and ENAs]{Adiabatic energy change in the inner heliosheath: How does it affect the distribution of pickup protons and energetic neutral atom fluxes?}

\author[I. I. Baliukin et al.]{
I. I. Baliukin,$^{1,2,3}$\thanks{E-mail: igor.baliukin@gmail.com}
V. V. Izmodenov$^{1,2,4}$
and D. B. Alexashov$^{1,4}$
\\
$^{1}$Space Research Institute of Russian Academy of Sciences, Profsoyuznaya Str. 84/32, Moscow, 117997, Russia\\
$^{2}$Lomonosov Moscow State University, Moscow Center for Fundamental and Applied Mathematics, GSP-1, Leninskie Gory, Moscow, 119991, Russia\\
$^{3}$HSE University, 20 Myasnitskaya Ulitsa, Moscow 101000, Russia\\
$^{4}$Institute for Problems in Mechanics, Vernadskogo 101-1, Moscow, 119526, Russia
}

\date{Accepted XXX. Received YYY; in original form ZZZ}

\pubyear{2023}

\begin{document}
\label{firstpage}
\pagerange{\pageref{firstpage}--\pageref{lastpage}}
\maketitle

\begin{abstract}

The hydrogen atoms penetrate the heliosphere from the local interstellar medium, and while being ionized, they form the population of pickup protons. The distribution of pickup protons is modified by the adiabatic heating (cooling) induced by the solar wind plasma compression (expansion). In this study, we emphasize the importance of the adiabatic energy change in the inner heliosheath that is usually either neglected or considered improperly. The effect of this process on the energy and spatial distributions of pickup protons and energetic neutral atoms (ENAs), which originate in the charge exchange of pickup protons, has been investigated and quantified using a kinetic model. The model employs the global distributions of plasma and hydrogen atoms in the heliosphere from the simulations of a kinetic-magnetohydrodynamic model of solar wind interaction with the local interstellar medium. The findings indicate that the adiabatic energy change is responsible for the broadening of the pickup proton velocity distribution and the significant enhancement of ENA fluxes (up to $\sim$5 and $\sim$20 times in the upwind and downwind directions at energies $\sim$1--2 keV for an observer at 1 au). It sheds light on the role of adiabatic energy change in explaining the discrepancies between the ENA flux observations and the results of numerical simulations.

\end{abstract}




\section{Introduction} \label{sec:intro}

The local interstellar medium (LISM) is partially ionized, and the interstellar neutral atoms penetrate the heliosphere due to the relative motion of the Sun and LISM with velocity of $\sim$26 km s$^{-1}$ \citep[e.g.][]{witte2004, mccomas2015}. The hydrogen atoms (the main neutral component of the LISM by its cosmic abundance) can be ionized owning to the processes of charge exchange with protons, photoionization, and electron impact. In the heliosphere, these newly born protons are picked up by the heliospheric magnetic field, forming the suprathermal component of protons (so-called pickup protons). The pickup protons are co-moving with the solar wind (SW) plasma, and their velocity distribution is determined by various processes such as the pitch-angle scattering, adiabatic heating/cooling, interaction with the heliospheric termination shock, stochastic acceleration induced by SW turbulence, and acceleration at propagating interplanetary shocks. \citet{sokol2022} provides an overview of present-day theory and modeling of pickup ions, while \citet{zirnstein2022} reviews existing in situ measurements.

In the inner heliosheath (IHS), the region between the heliospheric termination shock (TS) and the heliopause (HP), the solar wind plasma is slowed down and strongly heated (with temperature $\sim$10$^6$ K). There, the pickup protons can experience charge exchange with interstellar hydrogen atoms, creating energetic neutral atoms (ENAs) with energies of several keVs. ENAs have a large mean free path with respect to charge exchange \citep{izmod_etal2000} and carry significant information on the physical state of the region of their creation. The fluxes of ENAs are observed by different space based instruments in the vicinity of the Sun such as IBEX-Lo \citep[0.01--2 keV,][]{fuselier2009}, IBEX-Hi \citep[0.3--6 keV,][]{funsten2009}, Ion and Neutral Camera \citep[INCA, 5.2--55 keV,][]{krimigis2009}, and High-Energy Suprathermal Time-of-Flight sensor \citep[HSTOF, 58--88 keV, e.g.][]{hilchenbach1998} on board IBEX, Cassini, and Solar and Heliospheric Observatory (SOHO) spacecraft, respectively \citep[see also recent reviews by][]{galli2022, dialynas2022}. Besides direct observations made by Voyager 1 and 2, these data are the main source of knowledge on the inner heliosheath properties.

Applying numerical models to simulate the pickup proton distribution in the heliosphere and ENA fluxes from the inner heliosheath (so-called "globally distributed flux") has proven to be a powerful tool for data analysis. However, to make correct qualitative and quantitative conclusions based on the numerical simulations, the model should carefully treat all the important physical processes affecting the dynamics of the pickup protons. One of them is the adiabatic heating/cooling in the inner heliosheath that is often neglected \citep[e.g.][]{fahr_lay2000, zirnstein2015, zirnstein2016} or considered implicitly and improperly in the frame of fluid-type models that assume constant density and temperature fractions of the pickup protons throughout the IHS \citep[e.g.][]{zirnstein2017, kornbleuth2020, gkioulidou2022}, which is not physically justified. A distinctive feature of the model by \citet{baliukin2020, baliukin2022} is a rigorous kinetic treatment for pickup protons' distribution. This approach, among other things, allows taking into account the adiabatic heating (cooling) induced by the compression (expansion) of the solar wind plasma flow. 

In this paper, we emphasize the importance of the adiabatic heating/cooling in the inner heliosheath and quantify its effect on the velocity distribution of pickup protons and ENA fluxes using the kinetic model of the pickup proton distribution in the heliosphere by \citet{baliukin2020, baliukin2022}. The study is performed based on the plasma and neutral distributions obtained in the frame of the kinetic-magnetohydrodynamic (kinetic-MHD) model of the SW/LISM interaction developed by \citet{izmod2015, izmod2020}. Section \ref{sec:model} describes the model and methodology. In Section \ref{sec:results}, the main results of the work are presented. Section \ref{sec:conclusions} provides a summary along with a discussion.

\section{Model} \label{sec:model}

\subsection{Governing equations}

The kinetic equation for isotropic velocity distribution function of pickup protons $f_{\rm pui}^*(t, \mathbf{r}, w)$ in the SW plasma reference frame can be written in the following form (neglecting spatial and energy diffusion):
\begin{equation}
\frac{\partial f_{\rm pui}^*}{\partial t} + \mathbf{V} \cdot \frac{\partial f_{\rm pui}^*}{\partial \mathbf{r}} - \frac{w}{3} \frac{\partial f_{\rm pui}^*}{\partial w} {\rm div}(\mathbf{V}) = S_{+} - S_{-} f_{\rm pui}^*,
\label{eq:fpui_kinetic}
\end{equation}
where  $\mathbf{V}$ is the plasma bulk velocity, $w$ is pickup velocity in the SW reference frame, $S_{\rm +}$ and $S_{\rm -}$ are source and loss terms (in order to not overcharge the paper with expressions, we refer to \citet{baliukin2020}, see its equations 4 and 5). 

The third term on the left of equation (\ref{eq:fpui_kinetic}) is responsible for the change of velocity distribution function due to the adiabatic heating/cooling. We recognize the alternative theory of a pure magnetic energy change developed by \citet{fahr2007, fahr2011}. It is based on the conservation of particle invariants while being convected in an interplanetary magnetic field, which changes in magnitude. The application of this theory yields a different form of the third term \citep[see, e.g.,][]{fahr2016}. Interestingly, the magnetic cooling theory provides the $w^{-5}$ power-law shape of the pickup proton velocity distribution to speeds below the injection speed in the supersonic solar wind. Remarkable that the suprathermal tails in pickup proton distribution with the same spectral index of $-5$ were found by \citet{fisk2007}. However, in this work, we follow the classical theory with the adiabatic energy change term, leaving the study of the magnetic cooling effects for future works.

The formal solution of the kinetic equation (\ref{eq:fpui_kinetic}) with the boundary condition downstream of the TS is
\begin{align}
&f_{\rm pui}^{*}(t, \mathbf{r}, w) =  f_{\rm pui,d}^{*}(t_{\rm TS}, \mathbf{r}_{\rm TS}, w_{\rm TS}) \exp\left(- \int^t_{t_{\rm TS}} S_-(\hat{\tau}, \mathbf{r}(\hat{\tau}), w(\hat{\tau})) {\rm d} \hat{\tau} \right) \nonumber \\
&+ \int^t_{t_{\rm TS}} S_+(\tau, \mathbf{r}(\tau), w(\tau)) \exp\left(- \int^t_{\tau} S_-(\hat{\tau}, \mathbf{r}(\hat{\tau}), w(\hat{\tau})) {\rm d} \hat{\tau} \right) {\rm d} \tau, 
\label{eq:solution}
\end{align}
where $f_{\rm pui}^{*}$ and $f_{\rm pui,d}^{*}$ are the values of velocity distribution function at the particular moment $t$ and point $\mathbf{r}$ in the IHS and downstream of the TS, respectively, and subscript <<TS>> refers to the values of the parameters at the TS crossing. In this paper, we employ the power-law tail scenario for pickup proton velocity distribution function downstream of the TS, so $f_{\rm pui,d}^{*}$ is assumed to be the sum of the filled shell and the power-law tail ($f^*_{\rm tail} \propto w^{-\eta}$) distributions \citep[for details see section 2.1 in][]{baliukin2022}.

The integrations in equation (\ref{eq:solution}) are performed along the streamline (${\rm d} \mathbf{r}/{\rm d} t = \mathbf{V}$) with velocity change according to
\begin{equation}
\frac{{\rm d} w}{{\rm d} t} = -\frac{w}{3} {\rm div}(\mathbf{V}).
\label{eq:dwdt}
\end{equation}
Therefore, with ${\rm div}(\mathbf{V}) < 0$ particles experience adiabatic heating, which is the case for most of the inner heliosheath (as will be shown further). In turn, adiabatic cooling is operative in the spherically expanding supersonic SW, where 
\begin{equation}
{\rm div}(\mathbf{V}) = \frac{1}{r^2} \frac{{\rm d} (r^2 V)}{{\rm d} r} = \frac{{\rm d}V}{{\rm d} r} + \frac{2 V}{r} \approx \frac{2 V}{r} > 0.
\label{eq:div_sph}
\end{equation}

The equation (\ref{eq:dwdt}) can be integrated along the streamline, resulting in the relation between $w = w(t)$ and pickup proton velocity $w(t_{\rm TS})$ at the TS:
\begin{equation}
\frac{w}{w_{\rm TS}} = \exp \left( -\frac{1}{3} \int^t_{t_{\rm TS}}  {\rm div}(\mathbf{V}) {\rm d}t \right).
\label{eq:heating_factor}
\end{equation}
Hereafter, we refer to this ratio as an adiabatic heating factor, which indicates the change of the pickup proton velocity along the streamline starting from the TS. 

Although only stationary solutions will be sought in the paper, we prefer to keep time and terms with time derivatives in all equations to show their general mathematical structure.

\begin{figure*}
\includegraphics[width=\textwidth]{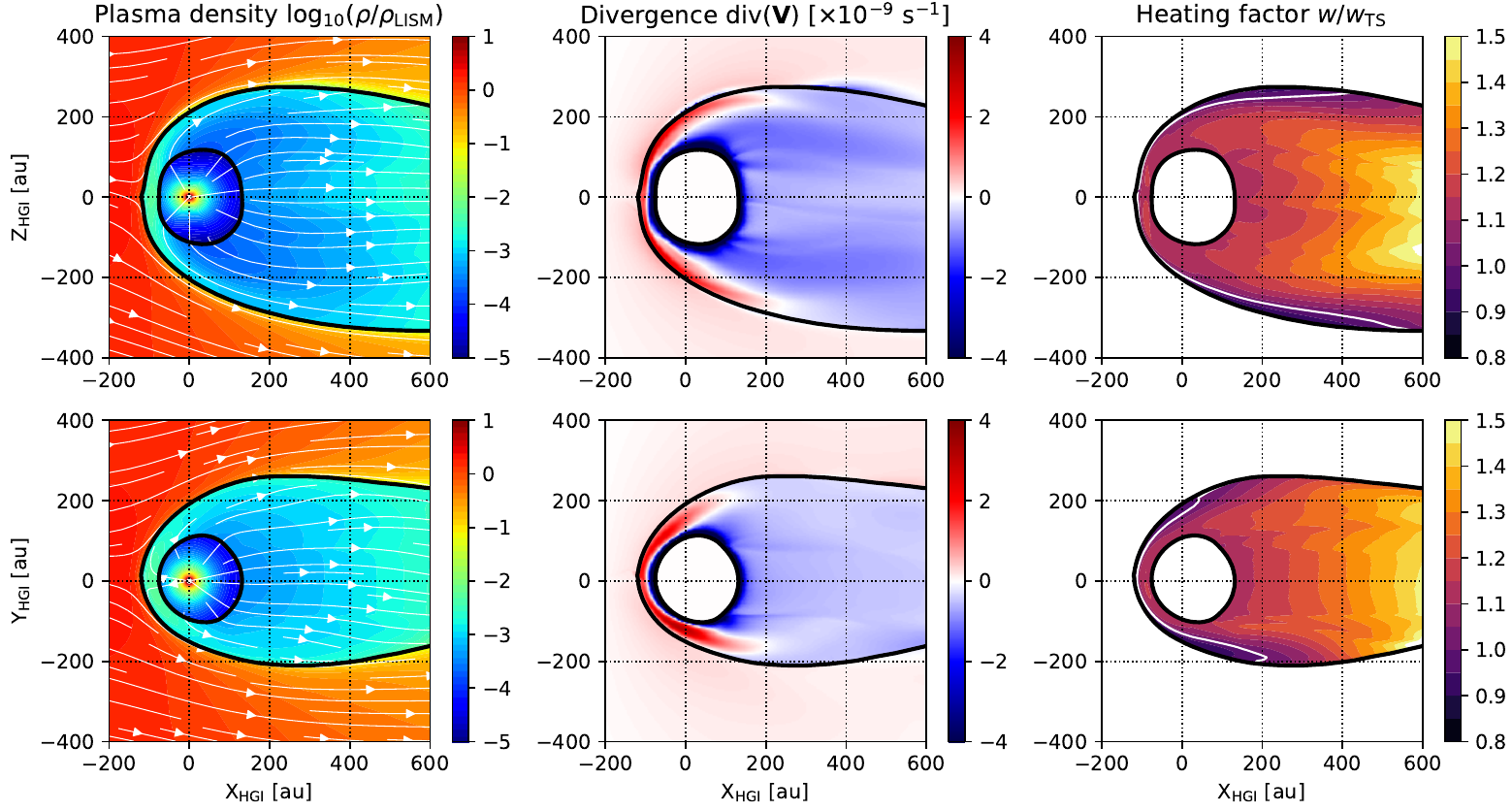}
\caption{
Meridional (top row) and solar equatorial (bottom row) slices of the heliosphere (in the heliographic inertial coordinate system). The first column shows the decimal logarithm $\log_{10}(\rho / \rho_{\rm LISM})$ of the plasma density (normalized by its LISM value) with superimposed plasma streamlines projected onto the corresponding plane (shown in white), second -- the divergence ${\rm div}(\mathbf{V})$ of the plasma bulk velocity, and third -- the heating factor $w / w_{\rm TS}$ (defined in the text). The divergence inside the TS is not shown for better representation since it is positive in the supersonic SW, and its absolute values are high. The white lines on the right panels show contours where the heating factor equals 1.0.
}
\label{fig:slices}
\end{figure*}

\subsection{Computation of divergence through the continuity equation}

The computation of the velocity divergence field is a non-trivial task since its numerical values depend on the topology and size of the computational cells. However, it is possible to account for the adiabatic energy change highly accurately by integrating the continuity equation along streamlines. This method does not require a direct numerical computation of derivatives of the velocity vector components.

The continuity equation for the plasma can be written in the following form:
\begin{equation}
\frac{{\rm d} \rho}{{\rm d} t} + \rho {\rm div}(\mathbf{V}) = q_1,
\end{equation}
where $\rho$ is the plasma density, and $q_1$ is the mass source, and ${\rm d} \rho/{\rm d} t$ is the derivative of density along the plasma streamline. 

In the global model of heliosphere by \citet{izmod2015, izmod2020} utilized in this work, the mass source is caused by the photoionization only (the electron impact ionization is not taken into account in the model). Therefore, $q_1 = m_{\rm p} n_{\rm H} \nu_{\rm ph}$, where $m_{\rm p}$ is  proton mass, $n_{\rm H}$ is the local hydrogen number density, and $\nu_{\rm ph}$ is the photoionization rate, which decreases $\propto 1/r^2$ with distance from the Sun. The divergence of the plasma bulk velocity is
\begin{equation}
{\rm div}(\mathbf{V}) = -\frac{1}{\rho} \frac{{\rm d} \rho}{{\rm d} t} + \frac{q_1}{\rho}.
\label{eq:div}
\end{equation}
The positive value of divergence characterizes the expansion and the negative -- compression of the solar wind plasma flow. In the supersonic solar wind ${\rm d}\rho / {\rm d}t < 0$ and $ {\rm div}(\mathbf{V}) > 0$.

After the substitution of equation (\ref{eq:div}) to equation (\ref{eq:heating_factor}) and integration, the following expression for the heating factor can be obtained:
\begin{equation}
\frac{w}{w_{\rm TS}} = \left( \frac{\rho}{\rho_{\rm TS}} \right)^{1/3} \cdot \exp \left( -\frac{1}{3} \int^t_{t_{\rm TS}} \frac{q_1}{\rho} {\rm d}t \right).
\label{eq:heating_factor2}
\end{equation}
The exponent factor given by mass sources suppresses the effect of adiabatic heating. Important to note that even though the photoionization rate in the inner heliosheath is low, it should not be omitted \citep[like it was done in, e.g.,][]{zirnstein2020}. Close to the HP in the nose region, where the stagnation of the flow occurs, this exponent is $\sim$0.95, according to our calculations. The accounting for the electron impact ionization, which may be especially effective in the IHS \citep{gruntman2015, chalov2019}, will lead to an even smaller value of the exponent, so additional attention should be paid to the mass source factor.

\subsection{Distribution from the global model of heliosphere}

The distributions of plasma and hydrogen atoms in the heliosphere have been calculated using the global kinetic-MHD model of the SW interaction with the LISM by \citet{izmod2020}. The main advantage of this model is the use of a moving computational grid and exact fitting of discontinuities -- the heliospheric termination shock in the solar wind and the heliopause separating the SW plasma from the interstellar plasma.

Figure \ref{fig:slices} shows meridional (top row) and solar equatorial (bottom row) slices of the heliosphere. The first column of this figure presents the decimal logarithm $\log_{10}(\rho / \rho_{\rm LISM})$ of the plasma density (normalized by its LISM value) with superimposed plasma streamlines projected onto the corresponding plane (shown in white), second -- divergence of the plasma bulk velocity ${\rm div}(\mathbf{V})$ calculated using equation (\ref{eq:div}), and third -- the heating factor $w / w_{\rm TS}$ defined by equation (\ref{eq:heating_factor2}). The divergence inside the TS is not shown for better representation since it is positive in the supersonic SW, and its absolute values are high.

As seen from Figure \ref{fig:slices}, right after the TS, plasma velocity divergence is negative. In the downwind hemisphere of the IHS, the plasma is predominantly compressed (the divergence is negative), and, accordingly, the adiabatic heating is operative. In the upwind hemisphere of the IHS, the divergence is generally positive. It is associated with the fact that after compression downstream of the TS in the nose region, the plasma finds a way to evacuate to other latitudes and longitudes, and the flow expands.

The heating factor shown in the third column of Figure \ref{fig:slices} is higher than 1.0 almost everywhere in the IHS (except the thin layers close to the HP). In the heliospheric tail, the heating factor increases to $\sim$1.5 at distances $\sim$600 au from the Sun, accompanied by the gradual increase of the plasma density (shown in the first column). Therefore, the effect of adiabatic heating is more pronounced in the heliotail.

\section{Results} \label{sec:results}
 
To study the effect of the adiabatic heating/cooling on the energy and spatial distributions of pickup protons and ENAs, the simulations were performed in two distinct cases -- with (Model 1) and without (Model 0) the adiabatic energy change in the IHS taken into account. In the calculations of Model 0, we explicitly assume that ${\rm div}(\mathbf{V}) = 0$ in the IHS, which is equivalent to setting the adiabatic heating factor $w / w_{\rm TS} = 1$.

\begin{figure}
\includegraphics[width=\columnwidth]{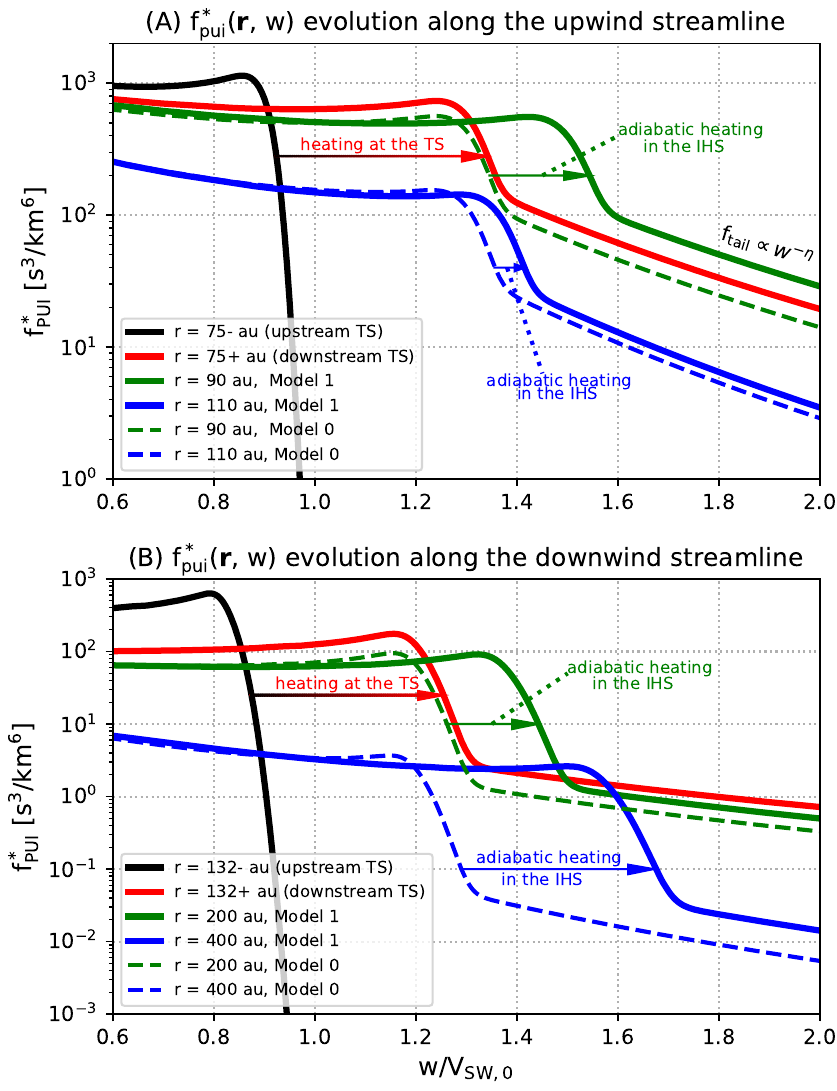}
\caption{
The distribution function $f^{*}_{\rm pui}$ as function of the velocity $w$ shown at different distances $r$ from the Sun (see the legend) along the upwind (panel A) and downwind (panel B) streamlines. Solid and dashed lines present the results of calculations using Model 1 and Model 0 (with ${\rm div}(\bmath{V}) = 0$ in the IHS assumed), respectively. $V_{\rm sw,0}$ = 432 km s$^{-1}$. To convert  the ratio $w / V_{\rm sw,0}$ to energy in the plasma reference frame, the formula $E ({\rm keV}) = 0.974 \times (w / V_{\rm sw,0})^2$ should be used. 
\label{fig:fpui}
}
\end{figure}

Figure \ref{fig:fpui} shows the profiles of the velocity distribution function of pickup protons at different distances from the Sun along the upwind (panel A) and downwind (panel B) streamlines. Solid and dashed lines present the results of calculations of Model 1 and Model 0, respectively. The transition from black to red curve represents the heating due to the compression at the heliospheric TS. The comparison of the solid red, dashed green, and dashed blue lines shows the effect of the gradual extinction of pickup protons (due to the charge exchange with background H atoms) on their way within the IHS. 

From the comparison of solid and dashed lines in Figure \ref{fig:fpui}, it can be concluded that the adiabatic heating manifests in the broadening of the distribution function, which, in turn, leads to an increase in the number density and kinetic temperature of pickup protons. To be more specific, 
\begin{equation}
\frac{n_{\rm PUI, 1}}{n_{\rm PUI, 0}} \approx \left( \frac{w}{w_{\rm TS}} \right)^3,\: \frac{T_{\rm PUI, 1}}{T_{\rm PUI, 0}} \approx \left( \frac{w}{w_{\rm TS}} \right)^2,
\label{eq:moments}
\end{equation}
according to the definition of the velocity distribution function moments, where subscripts <<1>> and <<0>> denote the parameters in Model 1 and Model 0, respectively.

In the nose region of the heliosphere (see panel A of Figure \ref{fig:fpui}), the process of adiabatic heating is efficient at small distances from the TS. The streamline emerging in the upwind direction first crosses the region of plasma compression downstream of the TS (transition from the red to green curve). After that, the adiabatic cooling becomes operative due to passing through the region of SW plasma expansion near the heliopause, where ${\rm div}(\mathbf{V}) > 0$ (transition from the green to blue curve). In the heliospheric tail (panel B), the adiabatic heating is effective all along the IHS since ${\rm div}(\mathbf{V}) < 0$. At $\sim$400 au from the Sun in the tail direction, the broadening of the distribution function reaches $\sim$30 \% (i.e., the heating factor $w / w_{\rm TS} \approx 1.3$). 

\begin{figure*}
\includegraphics[width=\textwidth]{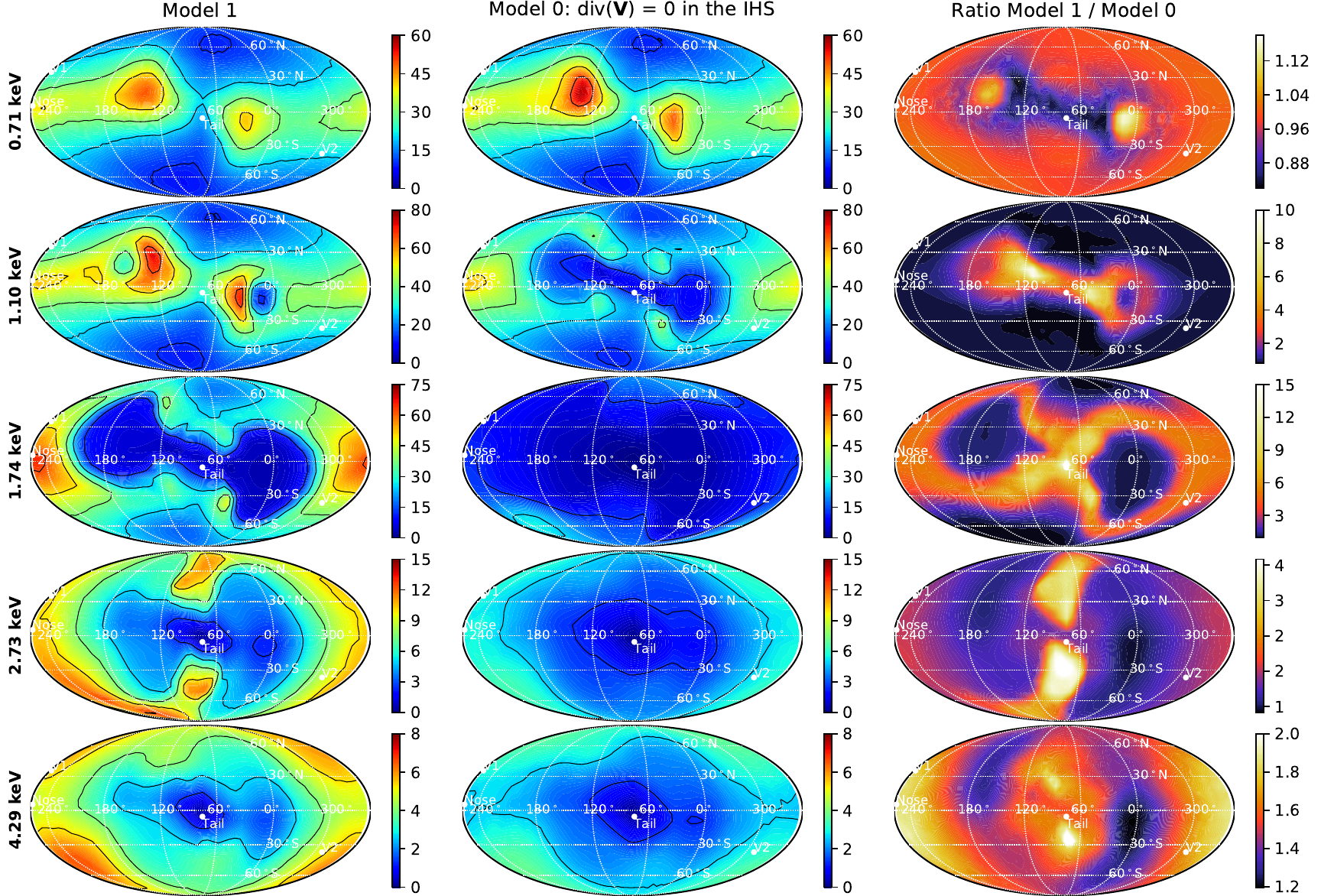}
\caption{
Full-sky maps (the Mollweide projections) of the globally distributed fluxes (ENAs originated in the IHS) in ecliptic (J2000) coordinates seen by the observer at 1 au at the energies 0.71, 1.1, 1.74, 2.73, and 4.29 keV, respectively (by rows). The first and second columns present the results of simulations in the frame of the Model 1 and Model 0 (with div($\mathbf{V}$) = 0 in the IHS assumed), respectively. The third column shows the ratio of ENA fluxes in Model 1 to Model 0. The units of fluxes are $(\rm cm^2\: sr\: s\: keV)^{-1}$. The maps are centered on the downwind longitude $75.4^\circ$ and $0^\circ$latitude.}
\label{fig:maps_nose}
\end{figure*}

Figure \ref{fig:maps_nose} shows the full-sky maps of the globally distributed fluxes (ENAs originated in the IHS) in ecliptic coordinates, as they are observed by the hypothetical instrument at energies 0.71, 1.1, 1.74, 2.73, and 4.29 keV from spacecraft at 1 au orbit around the Sun. The observational geometry and selected energy steps correspond to the capabilities of the IBEX-Hi instrument (and its energy channels 2--6) at Earth orbiting the IBEX spacecraft. We also note that for the sake of simplicity, the calculations were performed for the observer at rest, for the exact values of energy steps (without energy transmission taken into account), and the so-called Survival Probability correction was not applied to the simulated fluxes.

The first and second columns of Figure \ref{fig:maps_nose} present the results of simulations in the frame of Model 1 and Model 0, respectively, and the third column shows the ratio of ENA fluxes in Model 1 to Model 0. As can be seen, the adiabatic energy change is generally responsible for the flux increase (for most directions). The most prominent effect is seen in the heliotail, as also expected from the analysis of Figure \ref{fig:fpui}. At the lowest energy step (0.71 keV), the discrepancy between the models is minimal and does not exceed 20 \%. The biggest differences are seen in the intermediate energy steps (1.1 keV and 1.74 keV), and at the highest energy (2.73 keV and 4.29 keV), the ratio becomes smaller.

\begin{figure}
\includegraphics[width=\columnwidth]{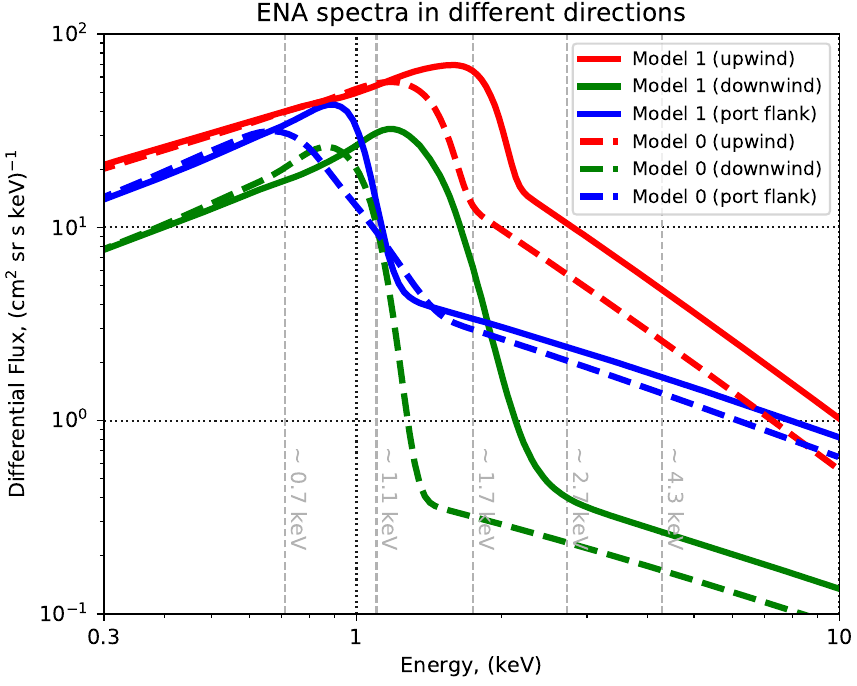}
\caption{
Spectra of globally distributed flux in different directions as seen by observed at 1 au. Red, green, and blue lines show the spectra in upwind, downwind, and port flank directions, respectively. Solid and dashed lines present the results of calculations in the frame of Model 1 and Model 0 (with div($\mathbf{V}$) = 0 in the IHS assumed), respectively. For the port flank direction, the ecliptic longitude 9$^\circ$ and ecliptic latitude -15$^\circ$ were used. 
\label{fig:spectra}
}
\end{figure}

To investigate the effect on the ENA fluxes in more detail, we have also calculated energy spectra of ENAs in some selected directions (see Figure \ref{fig:spectra}). Red, green, and blue lines show the spectra in upwind, downwind, and port flank directions, while solid and dashed lines correspond to calculations of Model 1 and Model 0, respectively. As expected, the strongest broadening of the spectrum is seen in the downwind direction (as for the velocity distribution function of pickup protons). In the upwind and downwind directions, the fluxes in Model 1 are generally higher than in Model 0. At 1.74 keV, Model 1 provides $\sim$5 and $\sim$20 times higher fluxes than Model 0 in the upwind and downwind directions, respectively. At the highest energy step (4.29 keV), the ratio stabilizes, and Model 1 fluxes are $\sim$1.8 and $\sim$1.6 times higher than in Model 0 in the upwind and downwind directions, respectively. Important to note that the exact values, however, depend on the shape of the velocity distribution function of pickup protons downstream of the TS (the power-law tail scenario was used in the calculations). 

In general, the steeper the spectral slope, the higher the (Model 1 / Model 0) ratio is. In the flank direction, the values of fluxes in the two models are comparable (see solid and dashed blue lines in Figure \ref{fig:spectra}). Interestingly, the adiabatic heating/cooling slightly modifies the spectral slope of the high-energy tail in the flank direction -- Model 1 provides a harder spectrum than Model 0.

\section{Summary and discussion} \label{sec:conclusions}

This study underlines the importance of the adiabatic energy change induced by the compression/expansion of the solar wind plasma in the inner heliosheath. We have investigated in detail the effect of this process on the energy and spatial distributions of pickup protons in the inner heliosheath and ENAs, and quantified it using the kinetic model developed by \citet{baliukin2020, baliukin2022}. The model employs the distributions of plasma and hydrogen atoms in the heliosphere from the simulations of the state-of-the-art kinetic-MHD model of the SW/LISM interaction. The main conclusions can be summarized as follows.

\begin{enumerate}

\item In the inner heliosheath, the plasma is predominantly compressed (except for the vicinity of the heliopause), and adiabatic heating is operative. To consider the adiabatic energy change under the kinetic description of pickup protons, the divergence of the plasma bulk velocity needs to be integrated along the streamlines. Instead of direct computation of the divergence, we derive it from the continuity equation. This approach allows improving the accuracy of calculations significantly.

\item The adiabatic heating manifests in the broadening of the velocity distribution function of pickup protons and, accordingly, in the number density and kinetic temperature increase. The influence of this process is pronounced the most in the tail region of the heliosphere, where the plasma is compressed. In the nose region of the inner heliosheath, adiabatic heating is less efficient since the corresponding streamlines cross both compression and expansion regions of the solar wind plasma.

\item The globally distributed fluxes are also strongly influenced by adiabatic heating/cooling. The simulations of the full-sky maps of ENA fluxes for a hypothetical observer at 1 au with IBEX-Hi capabilities show that the process under study is responsible for the substantial flux increase (up to $\sim$5 and $\sim$20 times in the upwind and downwind directions at energies $\sim$1--2 keV for an observer at 1 au). Therefore, for the correct interpretation of the data using models, the adiabatic energy change in the IHS must be taken into account.
\end{enumerate}

It should be noted that most studies on the globally distributed fluxes report the quantitative inconsistency between the results of numerical simulations and the data observed by IBEX-Hi and Cassini/INCA instruments. For example, \citet{gkioulidou2022} show that the data fluxes are systematically higher than the model ones (in the whole energy range covered by these instruments), and the magnitude of the difference depends on energy. In this regard, the authors speculate on the existence of an additional (currently unknown) process of acceleration of pickup protons in the inner heliosheath. \citet{kornbleuth2021} reported a deficit of model ENA fluxes with respect to the IBEX-Hi data, which is most notable in Voyager 1 and downwind directions (see figure 8 in their paper). The results of our work show that rigorous consideration of the adiabatic heating/cooling can explain (at least partially) the existing quantitative difference between the observations and results of numerical calculations.

In addition, there is a debate in the scientific community regarding the shape of the heliospheric tail \citep[see, e.g.][]{opher2015, izmod2015, kleimann2022}. The recent study by \citet{kornbleuth2023} suggests a promising possibility of resolving this question using the ENA flux data at high energies ($\sim$80 keV). However, one should be aware that conclusions on the heliotail shape based on ENA flux data and model simulations can be done only with proper consideration of the adiabatic energy change because, as our findings suggest, in the tail of the heliosphere, this process is very effective.

To conclude, we mention a recent work by \citet{wang2023}, where adiabatic energy change was also discussed. First, a crude oversimplification with a linear relationship between the plasma flow velocity and the distance from the termination shock was used in their work. Even though \citet{wang2023} came to qualitatively correct conclusions regarding the effect of the process on pickup protons and ENAs (for a single sky direction), the incorrect formula for the divergence of the plasma velocity was used: ${\rm div} (\mathbf{V}) = {\rm d}V/{\rm d} l$, where $V$ is the plasma bulk velocity, and $l$ is the coordinate along the streamline (see their equation 32). This formula does not account for curvilinearity, being correct only in the case of parallel flow. Such expression for the divergence describes only the first term of equality (\ref{eq:div_sph}), which, in turn, can be used in the region of the supersonic solar wind, and only in the nose and tail regions of the inner heliosheath, where the flow is nearly spherical. In the supersonic solar wind  ${\rm d}V/{\rm d} r \ll 2 V/r$, while right downstream of the termination shock, these terms are the same order of magnitude and have different signs. Therefore, the divergence expression above can be used neither in the supersonic solar wind nor the inner heliosheath, and the conclusions of their work can not be considered definitive.

\section*{Acknowledgements}
The work was performed in the frame of the Russian Science Foundation grant 19-12-00383.

\section*{Data availability}
The data underlying this article will be shared on reasonable request to the corresponding author.

\bibliographystyle{mnras}
\bibliography{mybib}




\bsp	
\label{lastpage}
\end{document}